\documentclass[11pt,a4paper]{article}
\pdfoutput=1
\RequirePackage{jcappub}

\title{\boldmath Taylor expansion of luminosity distance in Szekeres cosmological models: \\ Effects of local structures evolution on cosmographic parameters}

\author{Mattia Villani}
\affiliation{Sezione INFN di Firenze, \\Polo Scientifico Via Sansone 1, 50019, Sesto Fiorentino (FI), Italy }
\emailAdd{villani@fi.infn.it}

\abstract{We consider the Goode-Wainwright representation of the Szekeres cosmological models and calculate the Taylor expansion of the luminosity distance in order to study the effects of the inhomogeneities on cosmographic parameters.

Without making a particular choice for the arbitrary functions defining the metric, we Taylor expand up to the second order in redshift for Family I and up to the third order for Family II Szekeres metrics under the hypotesis, based on observation, that local structure formation is over. 

In a conservative fashion, we also allow for the existence of a non null cosmological constant.}

\keywords{Szekeres metrics, luminosity distance, dark energy}

\begin{document}
\maketitle
\flushbottom

\section{Introduction}

At the end of the nineties, it was proven that the Universe expansion is accelerating \cite{accel1,accel2}.

Assuming that General Relativity (GR) is \emph{the} theory of gravitation (valid at every scale, from the Solar System to galaxy clusters), in order to explain this phenomenon one needs a new form of energy with negative pressure able to contrast gravitational attraction which, otherwise, would slow down expansion.

The most successful model proposed in order to explain observations is $\Lambda$CDM: the Universe is assumed (at large scales) homogeneous and isotropic; it is described by a flat Friedmann-Lema\^itre-Robertson-Walker (FLRW) metric; along with baryonic matter it contains a component of cold dark matter (in order to explain spiral galaxies rotation curves and structures formation) and a component of dark energy described by the cosmological constant $\Lambda$. At present time, the dark energy amounts to about 68\% of the content of the Universe \cite{planck}.

Even if $\Lambda$CDM is in very good agreement with observations (see \cite{planck} for recent results), it has some flaws (see \cite{DElibro,mott}, for example): the cosmological constant is usually interpreted in terms of vacuum energy, but its value is way too small: it is known, in fact, that there is a difference of over 100 orders of magnitude between the theoretical prediction and the actual value of $\Lambda$; furthermore the acceleration starts at redshifts close to zero (the \emph{coincidence problem}) and it is very difficult to explain why in $\Lambda$CDM model.
%
%
%

In the literature, numerous different approaches have been proposed in order to explain the acceleration and overcome $\Lambda$CDM problems: some try to devise a new theory of gravity, like $f(R)$ theories (see for example \cite{capo}); other consider new forms of energy or fields that have the same effects as the cosmological constant.

Another approach is to consider the effects of inhomogeneities: it is evident that, at least at small scales, the universe is not homogeneous and isotropic, therefore it is natural to ask whether inhomogeneities can have some effect on the expansion. Among these approaches we mention the cosmological backreaction (see for example \cite{effetti1} - \cite{back8} for an incomplete list), and the use of non homogeneous metrics solution of the Einstein's equations (see \cite{void}-\cite{cv2} among the others and \cite{pk} and \cite{bmk} for reviews on the argument).

In the former, one studies the effects of local structure formation and evolution on the cosmic expansion. In the latter, one studies the effect of inhomogeneities on observations. 

A good review of all these approaches is \cite{DElibro}.

In this paper we follow the last approach and use Szekers' models, exact non homogeneous solution of the Einstein's equations with non rotational dust. In a conservative fashion, we consider a non null cosmological constant. 

In particular we work with the Goode-Wainwright representation, in which it is evident that Szekeres models describe non-linear perturbation of a FLWR background. 

Our aim is to calculate the Taylor expansion of luminosity distance of Szekeres models in order to study the effects of inhomogeneities and structure evolution on the deceleration and jerk parameters.


This paper is organized as follows: in section \ref{sec:szeker}, we briefly review the properties and the definition of Szekeres models and define the redshift; in section \ref{sec:DL} we calculate the Taylor expansion of the luminosity distance; in section \ref{sec:dir} we define our effective cosmographic parameters ans discuss their direction dependence; in section \ref{sec:obser} we compare our effective parameters with the observed ones and discuss the effects that local inhomogeneities and local structures evolution have on the acceleration; in the appendix \ref{app:rel}, we report some relations needed in our calculation; in the appendix \ref{app:cs}, we report the Christoffel symbols of the considered metrics; and in appendix \ref{sec:geod} we calculate null geodesics.

We assume $8 \pi G = c = 1$.

\section{Szekeres cosmological models}\label{sec:szeker}

Szekeres cosmological models are solutions of the Einstein's Field equations of the type:\footnote{Coordinates in \eqref{eq:def} are stereographic projection (\cite{k}-\cite{bmk}).}
\begin{equation}\label{eq:def}
ds^2 = dt^2 - e^{-2\, \alpha} \, dr^2 - e^{-2\, \beta} \, (dx^2 + dy^2)
\end{equation}
with irrotational dust as a source (see \cite{szek} - \cite{bmk}). One can also include a cosmological constant as a source (see for example \cite{pk} and \cite{szekl})\footnote{Szafron \cite{sza} (see also \cite{pk,k,bmk}) considered radiation too, but we won't consider that.}. In their most general form they have no killing vectors \cite{nosim} and are therefore inhomogeneous.\footnote{Particular choices for the arbitrary functions can, nevertheless, lead to models with symmetries: all Friedmann-Robertson-Walker and Lema\^itre-Tollman-Bondi models can be obtained from the Szekeres ones (see \cite{pk} on this topic).}

There are two families of this metric: the one with $\beta,_r\not=0$ (Family I) and the one with $\beta,_r=0$ (Family II). The function $\alpha$ always depends on all of the variables.

Goode and Wainwright gave a representation of the metric for the case $\Lambda = 0$ in which it is evident that the models describe non linear perturbations of the FLRW metrics. It also encompasses properties of both families \cite{GW}. The metric is given by:
\begin{equation}
ds^2 = dt^2 - S^2\left[ H^2\, W^2\, dr^2 + e^{2\nu} \, \left( dx^2 + dy^2 \right) \right].
\end{equation}
The function $S(t,r)$ is the solution of the equation:
\begin{equation}
\left( \dfrac{\dot{S}}{S} \right)^2 = \dfrac{2\mathcal{M}}{S^3} - \dfrac{\kappa}{S^2} \qquad \kappa = \{ 0, \pm 1 \},
\end{equation}
where $\mathcal{M}(r)$ is an arbitrary function linked to the matter density (see later); in case a cosmological constant is present, we have \cite{szekl}:
\begin{equation}
\left( \dfrac{\dot{S}}{S} \right)^2 =  \dfrac{2\mathcal{M}}{S^3} + \dfrac{\Lambda}{3} - \dfrac{\kappa}{S^2}, \qquad \kappa = \{ 0, \pm 1 \}.
\end{equation}

With or without a cosmological constant, we have:
\begin{equation}
\begin{split}
H(t,x,y,r) &= A(t,x,y,r) - F(t,r) =\\
&= A(t,x,y,r) - \left[ \beta_{+}(r)\, f_{+}(t) + \beta_{-}(r) \, f_{-}(t) \right],
\end{split}
\end{equation}
where $f_{\pm}(t)$ are independent solutions of the (Raychaudhuri) equation:\footnote{See \cite{pk} for their analytical expression in case there is no cosmological constant, and \cite{szekl} for the case $\kappa=0$ and $\Lambda \not= 0$.}
\begin{equation}\label{eq:raych}
\ddot{F} + 2 \dfrac{\dot{S}}{S}\, \dot{F} - \dfrac{3 \mathcal{M}}{S^3} \, F = 0
\end{equation}
and the $\beta_{\pm}(r)$ functions are arbitrary: the choice $\beta_{\pm}\equiv0$ leads to Robertson-Walker models, written in an unusual coordinate system (see \cite{pk}).

The two families are defined as follows:
\subsection{Family I $(\beta,_r \not= 0)$}
$\mathcal{M},_r^2 + T,_r^2 \not= 0$, and $S = S(t,r)$.\footnote{$T(r)$ is the, position dependent, time of the big bang, see \cite{pk}.} Moreover:

\begin{equation}
e^\nu = \dfrac{f(r)}{a(r) \, \left( x^2 + y^2\right) + 2 \, b(r) \, x + 2 \, c(r) \, y + d(r)}
\end{equation}
$f$ is arbitrary and:
\begin{eqnarray} 
\label{eq:condition}
ad - b^2 - c^2 &=& \dfrac{\epsilon}{4}, \qquad \epsilon = \{ 0, \pm 1 \};  \\ \nonumber
\\
W^2 &=& (\epsilon - \kappa f^2)^{-1}; \\ \nonumber
\\
\beta_{+} &=& - \dfrac{\kappa f \mathcal{M},_r}{3\, \mathcal{M}}, \qquad \beta_{-} = \dfrac{f \, T,_r}{6\mathcal{M}};  \\ \nonumber
\\
A &=& f \, \nu,_r - \kappa\, \beta_{+}.
\end{eqnarray}

Functions $f$, $a$, $b$, $c$, $d$, $\mathcal{M}$ and $T$ are arbitrary, but one should remember that there is the freedom to make a change of coordinates of the form $r \mapsto g(r)$, so, keeping in mind condition \eqref{eq:condition}, there are actually five arbitrary functions (see also \cite{GW}).

\subsection{Family II $(\beta,_r = 0)$}

$\mathcal{M}$ and $T$ are constants, $S = S(t)$ and $W=1$. Moreover:
\begin{equation}
e^\nu = \dfrac{1}{1 + \dfrac{\kappa}{4} \, \left( x^2 + y^2 \right)} \qquad \kappa= \{0; \pm 1\}
\end{equation}

\begin{equation}
A = \left\{ \begin{array}{lr}
e^\nu \left\{ a(r) \, \left[ 1- \dfrac{\kappa}{4} \, \left( x^2 + y^2 \right) \right] + b(r) \, x + c(r) \, y\right\} - \kappa \, \beta_{+}(r) & \phantom{spazio} \kappa = \pm 1; \\ \nonumber
\\
a(r) + b(r) \, x + c(r) \, y + \dfrac{\beta_{+}(r)}{2} \, \left( x^2 + y^2 \right)  &\phantom{spazio} \kappa=0;
\end{array} \right.
\end{equation}
 $a, b, c$ are arbitrary. In \cite{szekl} it is shown that for the case $\kappa= 0$ and $\Lambda\not=0$, we have:
\begin{equation}
A = a(r) + b(r) \, x + c(r) \, y +\, \dfrac{\beta_{+}}{4} \, (x^2 + y^2) \, \sqrt[3]{4\, \mathcal{M}^2\, \Lambda} 
\end{equation}

Function $a$, $b$, $c$, $\beta_{\pm}$ are arbitrary, but, with the coordinate freedom reported above, one can see that there are actually four arbitrary functions (see \cite{GW}).

\subsection{Density contrast and shear}\label{sec:shear}

In both families the density is given by:
\begin{equation}
\rho(t,x,y,r) = \dfrac{6\mathcal{M}}{S^3}\, \dfrac{A}{H} = \dfrac{6\mathcal{M}}{S^3} \, \left( 1+ \dfrac{F}{H} \right).
\end{equation}
The density contrast is given by:
\begin{equation}
\delta := \dfrac{F}{H}
\end{equation}
for both families (but see \cite{LSSgrowth}). Since the density is non negative, one needs $\delta>-1$.

The non-null components of the observer's rate of shear tensor are given by \cite{struct}:
\begin{align}\label{eq:shear}
\sigma_{r}^r &= \dfrac{2}{3}\,\dfrac{\dot{\delta}}{1+\delta}:=\sigma; \\
\sigma_{x}^x &= \sigma_{y}^y = -\dfrac{1}{2}\, \sigma,
\end{align}
where we defined $\sigma := \dfrac{2}{3}\, \dfrac{\dot{\delta}}{1+\delta}$ the first eigenvalue of the rate of the shear tensor. 

We emphasize that it depends on $\dot{\delta}$, the rate of structure growth.



%

\subsection{Cosmographic parameters}

We define the Hubble parameter in the usual fashion as:\footnote{This is valid for both families, but we remind that $\mathcal{H}=\mathcal{H}(r,t)$ for Family I and $\mathcal{H}=\mathcal{H}(t)$ for Family II.}
\begin{equation}
\begin{split}
\mathcal{H} &= \dfrac{\dot{S}}{S} = \sqrt{\dfrac{2\, \mathcal{M}}{S^3} + \dfrac{\Lambda}{3} - \dfrac{\kappa}{S^2}}\\
&= \mathcal{H}_0 \, \sqrt{\dfrac{\Omega_m}{S^3} + \Omega_{\Lambda} + \dfrac{\Omega_{\kappa}}{S^2}}
\end{split}
\end{equation}
where we defined:
\begin{equation}
\Omega_m = \dfrac{6\, \mathcal{M}}{3 \, \mathcal{H}_O^2}, \qquad \Omega_\Lambda = \dfrac{\Lambda}{3 \, \mathcal{H}_O^2}, \qquad \Omega_{\kappa} = - \dfrac{\kappa}{\mathcal{H}_O^2}.
\end{equation}
It is easy to prove that we have also the usual relation for the present time cosmological parameters:
\begin{equation}
1 = \Omega_m + \Omega_{\Lambda} + \Omega_{\kappa}
\end{equation}

It is not difficult to prove that, if we define the deceleration parameter associated to $\mathcal{H}$ as:
\begin{equation}
q = - \dfrac{\ddot{S}}{S} \, \dfrac{1}{\mathcal{H}^2}
\end{equation}
we have, at the present epoch ($S\equiv1$):
\begin{equation}\label{eq:qo}
q_O = \dfrac{\Omega_m}{2} - \Omega_{\Lambda} = \dfrac{1 - \Omega_k}{2} - \dfrac{3}{2} \, \Omega_{\Lambda}.
\end{equation}

One can also define a \emph{jerk} parameter:
\begin{equation}
j = \dfrac{1}{\mathcal{H}^3} \, \dfrac{\dot{\ddot{S}}}{S}.
\end{equation}
It is easy to show that, at present time (with $S=1$), we have $j_O \equiv 1$.
\subsection{Null geodesics and redshift}\label{sec:null}

Null geodesics are defined by the condition:
\begin{equation}\label{eq:null}
\begin{split}
0 &= k^\alpha \, k_{\alpha}\\
0 &= (k^0)^2 - e^{2\nu}\, S^2 \, \left[ (k^1)^2 + (k^2)^2 \right] - (HS)^2 \, (k^3)^2
\end{split}
\end{equation}
where $k^\alpha = \dfrac{d x^\alpha}{d \lambda}$, and $\lambda$ is an affine parameter: $\lambda_O$ is the position of the observer, $\lambda_S$ is the position of the source. The geodesics equation is:
\begin{equation}
\dfrac{d\, k^\alpha}{d \lambda} = - \Gamma^\alpha_{\beta \gamma} \, k^\beta \, k^\gamma.
\end{equation}
We give their expression in appendix \ref{sec:geod} (the Christoffel symbols needed for the calculation are reported in appendix \ref{app:cs}).

The redshift $z$ is defined as:
\begin{equation} \label{eq:red}
1 + z = \dfrac{k^0(\lambda)}{k^0(\lambda_O)}.
\end{equation}

\section{Luminosity distance}\label{sec:DL}

Since there is no evidence for local structure growth, we make the assumption that $\dot{\delta}(0) = 0$; from equation \eqref{eq:shear}, this means that
\begin{equation}
\sigma(0) = 0.
\end{equation}

We base our work on Sachs equations \cite{DL}, and calculate angular diameter distance $D_A$ from \cite{bol}:
\begin{equation}
\left\{ \begin{array}{l}
\dfrac{d^2\, D_A}{d \lambda^2} = - \left( |\Sigma|^2 + \dfrac{1}{2} \, R_{\alpha\beta} \, k^\alpha\, k^\beta \right) \, D_A\\
\\
\dfrac{d\Sigma}{d\lambda} + 2 \, \left(\dfrac{d}{d\lambda} \ln{D_A}\right) \, \Sigma = C_{\alpha\beta\mu\nu}\, \epsilon^{*\alpha}\, k^\beta\, \epsilon^{*\mu} \, k^\nu
 \end{array} \right.
\end{equation}
where: $\Sigma$ is the shear of the light bundle, $R_{\alpha\beta}$ is the Ricci tensor, $C_{\alpha\beta\mu\nu}$ is the Weyl tensor, $k^\mu$ is the null vector, $\epsilon^\mu$ is space-like and orthogonal to $k^\mu$, and, finally, $\lambda$ is the affine parameter.

In terms of redshift, we have, for the first equation:
\begin{equation}\label{eq:def_da}
\left( \dfrac{dz}{d\lambda} \right)^2 \, \dfrac{d^2 \, D_A}{dz^2} + \left( \dfrac{d^2 z}{d\lambda^2} \right) \, \dfrac{d \, D_A}{dz} =  - \left( |\Sigma|^2 + \dfrac{1}{2} \, R_{\alpha\beta} \, k^\alpha\, k^\beta \right) \, D_A.
\end{equation}

The initial conditions are \cite{mott}:
\begin{equation}
D_A(0) = 0, \qquad \dfrac{d \, D_A}{dz} \Big|_O = \dfrac{1}{\mathcal{H}_O}.
\end{equation}
We also impose $\Sigma(0) = 0$ (see \cite{szekl}). 

The term depending on Ricci tensor, is given by (we use the definition of the redshift, eqn. \eqref{eq:red}):
\begin{equation}
\dfrac{1}{2} \, R_{\alpha\beta} \, k^\alpha\, k^\beta= \dfrac{3}{2} \, \Omega_m\, \left( 1+ \delta \right)\, \left( \mathcal{H} \, k^0\right)_O^2 \, \dfrac{(1+z)^2}{S^3}.
\end{equation}

The coefficients in front of derivatives can be calculated from the definition of redshift given in equation \eqref{eq:red} and the geodetic equations given in appendix \ref{sec:geod}:
\begin{align}
\label{eq:dzdl}
\left(\dfrac{dz}{d\lambda}\right)^2 &= \left(\mathcal{H} \, k^0(0)\right)^2 \, (1+z)^4 - 2 \, \left(\mathcal{H}k^0(0)\right) \, \left(\dfrac{3}{2} \, \dfrac{\sigma}{k^0(0)} \right) \, \left(k^3HWS\right)^2 +\\ \nonumber
&+ \left( \dfrac{3}{2} \, \dfrac{\sigma}{k^0(0)} \right)^2 \, \left(k^3HWS\right)^4, \\ \nonumber
\\
\label{eq:d2zd2l}
\dfrac{d^2z}{d\lambda^2} & = - \dfrac{d\mathcal{H}}{d\lambda} \, k^0(0) \, (1+z)^2 - 2(1+z) \, \left( \mathcal{H}k^0(0) \right) \, \dfrac{dz}{d\lambda} + \\ \nonumber
& + \left(\dfrac{3}{2} \, \dfrac{d\sigma}{dz} \, \dfrac{1}{k^0(0)} \right) \, \dfrac{dz}{d\lambda} \, \left( k^3HWS \right)^2 + \left( \dfrac{3}{2} \, \dfrac{\sigma}{k^0(0)} \right) \, \dfrac{d}{d\lambda} \left( k^3HWS \right)^2,
\end{align}
where we remind that $W\equiv1$ for Family II models and, using the definition of $k^\alpha$ in section given \ref{sec:null}:\footnote{In the following a dot $\dot{}$ stands for a time derivative, while a prime ${}^\prime$ stands for a redshift derivative.}
\begin{equation}\label{eq:dhdl}
\dfrac{d\mathcal{H}}{d\lambda} = \left\{ \begin{array}{lcl}
k^0 \, \dot{\mathcal{H}} + k^3 \, \dfrac{\partial \mathcal{H}}{\partial r} && \text{Fam I}; \\
\\
k^0 \, \dot{\mathcal{H}} && \text{Fam II}.
\end{array} \right.
\end{equation}
It is easy to show that in the first line of equations \eqref{eq:dhdl}, one has: 
\begin{equation}\label{eq:radial}
\dfrac{1}{\mathcal{H}_O} \, \left( \dfrac{\partial \, \mathcal{H}}{\partial r} \right)_O = \dfrac{1}{\mathcal{H}_O^2} \, \dfrac{\partial \mathcal{M}}{\partial r} \Bigg|_O - \dfrac{\partial S}{\partial r} \Bigg|_O \, \left( q_O + 1 \right);
\end{equation}
so, in order to have a definite expression for this term, one has to give an expression at least for the arbitrary functions $\mathcal{M}(r)$ and $T(r)$, since these are the arbitrary functions defining $S$ (see section \ref{sec:szeker} and \cite{pk}).

We are interested in the effects of the inhomogeneities on the acceleration of the expansion and on the possible effects on the jerk parameter (which contains information on the dark energy equation of state). Therefore we only need the Taylor expansion of $D_A(z)$ up to the third order:
\begin{equation}
D_A(z) = \dfrac{z}{\mathcal{H}_O} + \dfrac{z^2}{2 \, \mathcal{H}_O} \, A + \dfrac{z^3}{6\, \mathcal{H}_O} \, B + O(z^4).
\end{equation}
We susbstitute it in \eqref{eq:def_da} with Taylor expanded coefficients \eqref{eq:dzdl} and \eqref{eq:d2zd2l}, in order to get $A$ e $B$.

In the following two subsection we will give an expression for $A$ and $B$ for the two families.

\subsection{Family I Szekeres cosmological models}

Since our aim here is not to construct a phenomenological model, we will not try to give a definite expression for the arbitrary functions, instead we will stop at the second order for Family I and study the effects of the inhomogenehities only on the deceleration parameter.\footnote{Without a definite expression for the arbitrary functions, we find a proliferation of terms like $\dfrac{S,r}{S}$ and $\dfrac{\dot{S},r}{S}$ difficult to manage.}

In Family II we do not need to specify a model and we will go up to the third order and study the effects of inhomogeneities on the jerk parameter.

It is easy to get:
\begin{equation}
A = - 3 - q_O + \left( \dfrac{3}{2} \, \dfrac{\sigma^\prime}{\mathcal{H}}\Bigg|_O \right) \, \left( \dfrac{k^3}{k^0} \right)^2_O + \left(\dfrac{1}{\mathcal{H}_O} \, \dfrac{\partial \mathcal{H}}{\partial r}\Bigg|_O \right) \, \left(\dfrac{k^3}{k^0}\right)_O,
\end{equation}
where $q_O$ is given in \eqref{eq:qo}.

Luminosity distance is given by:
\begin{align}
\label{eq:defdl}
D_L(z) &= (1+z)^2 \, D_A(z) \\
&= \dfrac{D_L^{(1)}}{\mathcal{H}_0} z + \dfrac{D_L^{(2)I}}{2 \, \mathcal{H}_0} \, z^2 + O(z^3),
\end{align}
where:
\begin{subequations}
\begin{equation}
D_L^{(1)} = 1;
\end{equation}
\begin{equation}
\begin{split}
D_L^{(2)I} &= 1 - q_O + \left( \dfrac{3}{2} \, \dfrac{\sigma^\prime}{\mathcal{H}}\Bigg|_O \right) \, \left( \dfrac{k^3}{k^0} \right)^2_O + \left(\dfrac{1}{\mathcal{H}_O} \, \dfrac{\partial \mathcal{H}}{\partial r}\Bigg|_O \right) \, \left(\dfrac{k^3}{k^0}\right)_O =\\
&= 1 - q_{eff}^{I},
\end{split}
\end{equation}
\end{subequations}
where we defined the effective deceleration parameter for Family I:
\begin{equation} \label{eq:deaccI}
q_{eff}^I := q_O - \left( \dfrac{3}{2} \, \dfrac{\sigma^\prime}{\mathcal{H}}\Bigg|_O \right) \, \left( \dfrac{k^3}{k^0} \right)^2_O - \left(\dfrac{1}{\mathcal{H}_O} \, \dfrac{\partial \mathcal{H}}{\partial r}\Bigg|_O \right) \, \left(\dfrac{k^3}{k^0}\right)_O.
\end{equation}
It depends on the radial derivative of the Hubble parameter given in \eqref{eq:radial} (the cosmic variance problem), on the direction of observation (see section \ref{sec:dir}) and on $\sigma^\prime(0)$, which, we remind, is linked to the inhomogeneities growth (see equation \eqref{eq:shear}).

\subsection{Family II Szekeres cosmological models}\label{sec:famII}

As already said, in the case of Family II cosmological models, we can easily Taylor expand the angular diameter distance up to the third order without specifying any particular model.

Using equations \eqref{eq:k0fam1}, \eqref{eq:k3fam1}, we find:
\begin{equation}
A = -3 - q_O + \left( \dfrac{3}{2} \, \dfrac{\sigma^\prime}{\mathcal{H}}\Bigg|_O \right) \, \left( \dfrac{k^3}{k^0} \right)^2_O,
\end{equation}
for the second order, while for the third order, we have:
\begin{equation}
\begin{split}
B &= 3\, A^2 - j_O - 10\, \left[ q_O - \left( \dfrac{3}{2}\, \dfrac{\sigma^\prime}{\mathcal{H}} \right) \, \left( \dfrac{k^3}{k^0} \right)^2 \right]_O -15 + 2 \, \left( \dfrac{3}{2} \, \dfrac{\sigma^\prime}{\mathcal{H}}\Big|_O \right) \, \left( \dfrac{k^3}{k^0} \right)_O^2+\\
& + 4 \, \left( \dfrac{3}{2} \, \dfrac{\sigma^\prime}{\mathcal{H}}\Big|_O \right) \, \left( \dfrac{k^3}{k^0} \right)_O^2 \, \left[ \dfrac{k^1}{k^0} \, \dfrac{\partial}{\partial x} \, \ln{H} + \dfrac{k^2}{k^0} \, \dfrac{\partial}{\partial y} \, \ln{H} \right]_O +\\
&- \dfrac{3}{2} \, \Omega_m \, \left( 1+\delta(0) \right).
\end{split}
\end{equation}

Luminosity distance is given, as usual, by $D_L(z) = (1+z)^2\, D_A(z)$, and we find:
\begin{equation}
\begin{split}
D_L(z)  &= \dfrac{z}{\mathcal{H}_O} \; \left( 1+2z+z^2\right)\; \left( 1 + \dfrac{z}{2} \, A + \dfrac{z^2}{6} \, B + O(z^3) \right) =\\
&=\dfrac{z}{\mathcal{H}_O} + \dfrac{z^2}{2 \, \mathcal{H}_O} \, \left( 4 + A \right) + \dfrac{z^3}{6 \, \mathcal{H}_O} \, \left( B + 6A +6 \right) + O(z^4)=\\
&= \dfrac{D^{(1)}_L}{\mathcal{H}_O}\, z + \dfrac{D^{(2)II}_L}{2\, \mathcal{H}_O}\, z^2 + \dfrac{D_L^{(3)II}}{6 \, \mathcal{H}_O}\, z^3 + O(z^4),
\end{split}
\end{equation}
where:
\begin{subequations}
\begin{equation}
D^{(1)}_L = 1;
\end{equation}
\begin{equation}
\begin{split}\label{eq:dl2}
D^{(2)II}_L &= 1 - q_O + \left( \dfrac{3}{2} \, \dfrac{\sigma^\prime}{\mathcal{H}}\Big|_O \right)\, \left( \dfrac{k^3}{k^0} \right)^2_O \\
&= 1 - q^{II}_{eff};
\end{split}
\end{equation}
\begin{equation}\label{eq:ter}
\begin{split}
D^{(3)II}_L &= 3 \, (q^{II}_{eff})^2 + q_{eff}^{II} - j_O - \Omega_O + \left(\dfrac{3}{2} \, \dfrac{\sigma^\prime}{\mathcal{H}}\Big|_O \right)+  \,\left( \dfrac{3}{2}\, \dfrac{\sigma^\prime}{\mathcal{H}}\Big|_O \right)\, \left( \dfrac{k^3}{k^0} \right)_O^2 + \\
&+ 4 \, \left( \dfrac{3}{2} \, \dfrac{\sigma^\prime}{\mathcal{H}} \Big|_O \right) \, \left( \dfrac{k^3}{k^0} \right)^2_O \, \left[ \dfrac{k^1}{k^0} \, \dfrac{\partial}{\partial x} \, \ln{H} + \dfrac{k^2}{k^0} \, \dfrac{\partial}{\partial y} \, \ln{H} \right]_O .
\end{split}
\end{equation}
\end{subequations}
In the previous equation, we defined the parameter $\Omega_O= 1 - \Omega_k$ (see also \cite{capo}), and the effective deceleration parameter for Family II models, $q_{eff}^{II}$:
\begin{equation}\label{eq:deaccII}
q_{eff}^{II} := q_O - \left( \dfrac{3}{2} \, \dfrac{\sigma^\prime}{\mathcal{H}} \Bigg|_O\right) \, \left( \dfrac{k^3}{k^0} \right)^2_O.
\end{equation}

\section{Angular dependence}\label{sec:dir}

The terms $\left( \dfrac{k^i}{k^0} \right)\Bigg|_O$ that appear in the previous equations are direction dependent. 

Following, for example, \cite{szekl}, we define, at the observer position, $\alpha$, the angle formed by the $k^\mu$ with the $r$ axis, and $\beta$, the angle formed with the $r-x$ plane;  in this way we have:
\begin{align}
\label{eq:angk1}
k^1(0) &= k^0(0) \, \cos{\beta}\, \sin{\alpha};\\
k^2(0) &= k^0(0)\, \sin{\beta} \, \sin{\alpha};\\
\label{eq:angk3}
k^3(0) &= k^0(0) \, \cos{\alpha}.
\end{align}

\subsection{The effective deceleration parameters}\label{sec:qeff}

The usual expression for the second order of the luminosity distance is \cite{weinberg}:
\begin{equation}
D^{(2)}_L = 1 - q_{obs},
\end{equation}
where $q_{obs}$ is the observed deceleration parameter.

We defined the deceleration parameters in equation \eqref{eq:deaccI} for Family I and in equation \eqref{eq:deaccII} for Family II. Using equations \eqref{eq:angk1}-\eqref{eq:angk3}, we rewrite them respectively as:
\begin{align}
q_{eff}^I &= q_O - \left(\dfrac{1}{\mathcal{H}_O} \, \dfrac{\partial \mathcal{H}}{\partial r}\Bigg|_O \right) \, \cos{\alpha} - \left( \dfrac{3}{2} \, \dfrac{\sigma^\prime}{\mathcal{H}}\Big|_O \right) \, \cos^2{\alpha} = \\ \nonumber
\\ \label{eq:qeffdel1}
&= q_O - \left(\dfrac{1}{\mathcal{H}_O} \, \dfrac{\partial \mathcal{H}}{\partial r}\Bigg|_O \right) \, \cos{\alpha} +  \dfrac{3}{2} \, \delta(0)\, \Omega_M \, \cos^2{\alpha};\\ \nonumber
\\
q_{eff}^{II} &= q_O - \left(\dfrac{3}{2} \, \dfrac{\sigma^\prime}{\mathcal{H}}\Bigg|_O \right)\, \cos^2{\alpha} = \\ \nonumber
\\  \label{eq:qeffdel2}
&= q_O +\dfrac{3}{2} \, \Omega_M\, \delta(0) \, \cos^2{\alpha}.
\end{align}
where in \eqref{eq:qeffdel1} and \eqref{eq:qeffdel2}, we used the relation:
\begin{equation}\label{eq:rel}
\dfrac{\sigma^\prime}{\mathcal{H}}\Bigg|_O = - \delta(0) \, \Omega_m 
\end{equation} 
derived in the appendix A.

We notice how in both cases the presence of a local inhomogeneity gives rise to an angular dependence of the deceleration parameter; in Family I we see that the dependence of the Hubble parameter on the radial coordinate, gives rise to an additional angular dependence. We notice further that only the angle $\alpha$, formed by the direction of observation and the $r$ axis, is involved in the expression.

The dependence of the Hubble parameter on the radial coordinate is a form of \emph{cosmic variance}: different observers at different positions measure different value for the cosmographic parameter. As already argued by \cite{cv} this might mitigate the tension between the local measurement of $\mathcal{H}$ and the ones coming from BAO and CMB\footnote{See \cite{cv1} and \cite{cv2} for older works on the same topic.}. 

Averaging over directions gives:
\begin{equation}
\begin{split}
\langle q_{eff}^I \rangle = \langle q_{eff}^{II} \rangle &= \dfrac{1 - \Omega_k}{2} - \dfrac{3}{2} \, \left[ \Omega_{\Lambda} + \dfrac{1}{3} \, \dfrac{\sigma^\prime}{\mathcal{H}}\Big|_O \right] = \\
&= \dfrac{1 - \Omega_k}{2} \, \left( 1 + \delta(0) \right) - \dfrac{3}{2} \, \Omega_{\Lambda} \, \left( 1 + \dfrac{\delta(0)}{3} \right).
\end{split}
\end{equation}
From the first equality we notice that if $\sigma^\prime (0) > 0 $ (and therefore $\delta(0) < 0 $, from \eqref{eq:rel1}), less dark energy is needed in order to explain acceleration.

\subsection{The effective jerk parameter $j_{eff}$}\label{sec:jeff}

The usual expression for $D_L^{(3)}$ in $\Lambda$CDM models (for arbitrary values of $\kappa$) is given by (see for example \cite{capo, pedi}):
\begin{equation}\label{eq:frw}
D^{(3)}_L = 3 q_{obs}^2 + q_{obs} -\Omega_O - j_{obs}
\end{equation}
where one expects $j_{obs} \equiv 1$ \cite{pedi,stf,de3}.

In our case we have, for Family II's third order:
\begin{equation}\label{eq:terang}
\begin{split}
D^{(3)II}_L &= 3 \, (q^{II}_{eff})^2 + q_{eff}^{II} - j_O - \Omega_O + \left(\dfrac{3}{2} \, \dfrac{\sigma^\prime}{\mathcal{H}}\Big|_O \right)+  \,\left( \dfrac{3}{2}\, \dfrac{\sigma^\prime}{\mathcal{H}}\Big|_O \right)\, \cos^2{\alpha} + \\
&+ 4 \, \left( \dfrac{3}{2} \, \dfrac{\sigma^\prime}{\mathcal{H}} \Big|_O \right) \, \cos^2{\alpha} \, \left[ \cos{\beta}\sin{\alpha} \, \dfrac{\partial}{\partial x} \, \ln{H} + \sin{\alpha}\sin{\beta} \, \dfrac{\partial}{\partial y} \, \ln{H} \right]_O
\end{split}
\end{equation} 

We can read off the expression for the effective jerk parameter by comparing the previous equation with \eqref{eq:frw}:
\begin{equation}\label{eq:jerk}
\begin{split}
 j_{eff} &=  j_O - \left(\dfrac{3}{2} \, \dfrac{\sigma^\prime}{\mathcal{H}}\Big|_O \right) - \,\left( \dfrac{3}{2}\, \dfrac{\sigma^\prime}{\mathcal{H}}\Big|_O \right)\, \cos^2{\alpha} + \\
&- 4 \, \left( \dfrac{3}{2} \, \dfrac{\sigma^\prime}{\mathcal{H}} \Big|_O \right) \, \cos^2{\alpha} \, \left[ \cos{\beta}\sin{\alpha} \, \dfrac{\partial}{\partial x} \, \ln{H} + \sin{\alpha}\sin{\beta} \, \dfrac{\partial}{\partial y} \, \ln{H} \right]_O=\\
&= j_O + \dfrac{3}{2} \, \Omega_M \,\delta(0) \, \left( 1 +  \cos^2{\alpha} \right)+ \\
& + 4 \, \left( \dfrac{3}{2} \, \Omega_M \, \delta(0) \right) \, \cos^2{\alpha} \, \left[ \cos{\beta}\sin{\alpha} \, \dfrac{\partial}{\partial x} \, \ln{H} + \sin{\alpha}\sin{\beta} \, \dfrac{\partial}{\partial y} \, \ln{H} \right]_O
\end{split}
\end{equation}
where, as we said earlier, $j_O \equiv 1$ and where we used equation \eqref{eq:rel1} in the second equatility. 

The presence of the inhomogeneity gives rise to a change of the value of the jerk from the $\Lambda$CDM one; moreover, in this case we see that there is a dependence on both angles $\alpha$ and $\beta$ coming form the $x$- and $y$-derivatives of $H$. From its definition, it is easy to see that those are the derivatives of the density contrast:
\begin{equation}
\dfrac{1}{H} \, \dfrac{\partial H}{\partial x} = - \dfrac{1}{\delta} \, \dfrac{\partial \delta}{\partial x} \qquad \dfrac{1}{H}\, \dfrac{\partial H}{\partial y} = - \dfrac{1}{\delta} \, \dfrac{\partial \delta}{\partial y}.
\end{equation}
Substituting in \eqref{eq:jerk}, we have:
\begin{equation}\label{eq:jeff}
\begin{split}
 j_{eff} &= j_O + \dfrac{3}{2} \, \Omega_M \,\delta(0) \, \left( 1 +  \cos^2{\alpha} \right)+ \\
& - 6 \, \Omega_M \, \cos^2{\alpha} \, \left[ \cos{\beta}\sin{\alpha} \; \dfrac{\partial \delta}{\partial x}  + \sin{\alpha}\sin{\beta} \; \dfrac{\partial \delta}{\partial y} \right]_O
\end{split}
\end{equation}

Averaging over directions, we have:
\begin{equation}
\langle j_{eff} \rangle = j_O + 2 \, \Omega_m \, \delta(0),
\end{equation}
therefore, $\delta(0)<0$ gives $\langle j_{eff} \rangle < 1$, a result found by some recent works, (see \cite{pedi}, \cite{capo}).

\section{Comparison with observations}\label{sec:obser}

As said in the introduction, the idea that inhomogeneities can have an effect on observation is not new. Many works in the literature are based on the assumption that we might live in a void (a region with $\delta < 0$), since a void expands at a faster rate than the the background, thus mimicking the effect of a cosmological constant.

In particular, in \cite{void2} the authors use 44 close SNe Ia (20-300$h^{-1}$ Mpc) to study the monopole of peculiar velocities of the host galaxies: they find that their data are consistent with a local void with an underdensity of about 20\%, so $\delta(0) \approx -0.20$.

In \cite{void5}, the authors fit 3-yr WMAP data and SNe Ia with the hypothesis that we live in a spherical-symmetric void described by a LTB metric: they find that they need $\delta(0) \approx -0.4$ and that the observer must be close to the center of the sphere for their results to be consistent with the CMB. 

More recently, in \cite{void12} the authors use a LTB void metric to describe a local void and fit their model using CMB, SN, BAO and $\mathcal{H}_0$ allowing for a non null space curvature: they find that their model fits the data with a $\delta(0)=-0.65$ void that extends up to $z\approx1$ (gigaparsec scale); they also find $\Omega_k = -0.2$ and that the model is consistent with the CMB dipole if the observer is displaced by about 15 Mpc from the center.

The fact that the observer must be close to the center of the void in order to be consistent with the CMB data is a violation of the Cosmological Principle, since that position is obviously a particular one. Even if the void is non-symmetric (so that there is not an actual center), there is still a weak violation of the Principle, since usually galaxies are found in clusters not in voids, see \cite{mott}.

In \cite{cosmop}, the authors propose a way to falsify non Copernican models by constraining the parameter space of those models; they give constraints on $\delta$ and on the radius of the inhomogeneity finding that only models with $\delta(0) \approx 0$ are admissible, though many values of the radius are allowed for the Copernican principle to be verified.

As far as the jerk parameter and the dark energy equation of state\footnote{See \cite{stf} for the link between the two.} is concerned, we see that inhomogeneities change the value from the expected $j_{obs}=$1: an observer might therefore interpret this effect as a change in the equation of state of dark energy with redshift.

A similar result is found also in \cite{uncw}: they consider a quasi-linear inhomogeneity\footnote{This inhomogeneity is constructed using the model developed in \cite{cold}.} in a flat $w$CDM background, with a constant $w$; they test their model with SNe, CMB and local measurement of the Hubble parameter and find $\delta(0) = 0.1 \div 0.15 $ and that $w=-1$ is excluded at 95\% of confidence level. They also find a degeneracy between $w$ and $\delta(0)$ and argue that even if $w$ is constant in the background, the effect of the inhomogeneities is to make it depend on the redshift for an observer that wanted to use a homogeneous model to fit the data. In \cite{uncw2}, the authors suggest that until the value of the local over- or under-density is known, their effect should be considered as a systematic error in $w$ measurements (systematic, because linked to the cosmic variance problem).

On this topic, see also \cite{starob}, where the authors use an LTB model with a cosmological constant in order to study the effect on the dark energy equation of state; they find that an underdesity leads to an apparent phantom behavoir of the dark energy, while an over density leads to an apparent quintessence behavoir.

\subsection{Effects of angular dependency}
From the discussion in the previous section, we have that an estimate for the value of the local density contrast is $\delta(0) \approx  -0.65 \div 0.15$. 

One can use the value of matter density as given by the Planck collaboration \cite{planck}, $\Omega_m = 0.135 \pm 0.017$, in order to estimate the corrections due to inhomogeneities: from Family II effective deceleration parameter, $q_{eff}^{II}$, these corrections are of percent level and should be measurable (indeed, see \cite{anis} for a claim of asymmetry in supernova data); in case of Family I, there is also the correction coming from the Hubble parameter radial derivative which amuonts to about 1 \%.

We can say nothing about the jerk parameter since, as far as we know, there are no measure for the density gradient, but the second term in equation \eqref{eq:jeff} gives again a correction at percent level.

\section{Conclusions}

Szekeres cosmological models are inhomogeneus and anisotropic solutions of the Einstein's Field equations with irrotational dust and a cosmological constant as a source.

Starting from the Sachs equations, we have calculated the Taylor expansion of the Szekeres luminosity distance for both Family I and Family II allowing for the existence of a cosmological constant. Relying on observations, we make the hypotesis that there is no structure growth at the observer position: $\dot{\delta}(0)=0$.

Confronting our luminosity distance with the usual expansion of the $\Lambda$CDM luminosity distance, we have found corrections proportional to $\sigma^\prime(0)$ that are also direction dependent. $\sigma$ is the eigenvalue of the observer's rate of the shear tensor $\sigma = \sigma_{r}^r$ and is linked to the evolution of the density contrast (see equation \eqref{eq:shear}). Therefore, inhomogeneities and local structure evolution affect the way an observer sees the expansion. 

In Family I models, at second order, there is also a correction proportional to $\dfrac{\partial\mathcal{H}}{\partial r}\Bigg|_O$, a consequence of the \emph{cosmic variance}. This could give an explanation to the tension between local direct extimates of $\mathcal{H}_O$ and indirect ones from BAO and CMB (see \cite{cv}-\cite{cv2}).

Using data from the literature, we find that inhomogeneities could give corrections of percent level on the deceleration parameter: direction dependence should therefore be observable. In the case of the effective jerk parameter, a measure of the density gradient is needed in order to give an estimate of the correction.

\appendix
\section{Useful relations}\label{app:rel}

In this appendix we give a derivation of equation \eqref{eq:rel}, which relates $\sigma^\prime(0)$ and the density contrast $\delta(0)$ at the observer's position, and we give an analogue expression also for $\sigma^{\prime\prime}(0)$.

Since, in Family II models, we can impose that $H_O = S_O = 1$, we have:
\begin{eqnarray}
\label{eq:sigmazero}
\dfrac{3}{2} \, \sigma(0) &=& \dfrac{\dot{F}}{H}\Bigg|_O = \dot{F}|_O = 0; \\  \nonumber
\\ \label{eq:dens}
\delta(0) &=& \dfrac{F}{H} \Bigg|_O = F|_O.
\end{eqnarray}
Since for Family I, we have $(HW)_O = 1$, we have, in this case:
\begin{equation}
\delta(0) = \dfrac{F}{H}\Bigg|_O = (FW)_O.
\end{equation}

Using the definition of redshift and of $k^0$ given in \ref{sec:null}, the first derivative of the $\sigma$ with respect to the redshift calculated at the position of the observer is:
\begin{equation}
\begin{split}
\dfrac{3}{2}\, \sigma^\prime(0) &= \dfrac{3}{2} \, \dfrac{d \sigma}{dz}\Big|_O = \\
\\
&=\dfrac{3}{2} \, \dfrac{d\lambda}{dz}\, \dfrac{dt}{d\lambda} \, \dfrac{d\sigma}{dt}\Big|_O=\\
\\
&= - \dfrac{3}{2} \, \dfrac{1}{\mathcal{H}_O} \, \dfrac{d\sigma}{dt}\Big|_O=\\
\\
&= - \dfrac{1}{\mathcal{H}_O} \, \left[ \dfrac{\ddot{F}}{H} + \left( \dfrac{\dot{F}}{H}\right)^2 \right]_O,
\end{split}
\end{equation}
where in the third line we used equation \eqref{eq:red} and the definition of $k^0$ given in \ref{sec:null}. In the last line, the second term is zero because of equation \eqref{eq:sigmazero}. We now use the Raychaudhury equation \eqref{eq:raych}, so we get:
\begin{equation}
\begin{split}
\dfrac{3}{2}\, \sigma^\prime(0) &= - \dfrac{1}{\mathcal{H}_O} \, \ddot{F}\Bigg|_O = -\dfrac{1}{\mathcal{H}_O} \, \left[ -2 \, \mathcal{H} \, \dot{F} + \dfrac{3\mathcal{M}}{S^3} \, F \right]_O =\\
\\
&= - \dfrac{1}{\mathcal{H}_O} \, \left[ \dfrac{3}{2} \, \Omega_m \, \mathcal{H}^2_O \, \delta(0) \right]
\end{split}
\end{equation}
In the second line we used again \eqref{eq:sigmazero} and \eqref{eq:dens} and also the definition of $\Omega_m$. Simplifying, we get:
\begin{equation} \label{eq:rel1}
\sigma^\prime(0) = - \Omega_m \, \delta(0) \, \mathcal{H}_O.
\end{equation}
which is the expression given in \eqref{eq:rel}.

On the same line, we find, for Family II:
\begin{equation}\label{eq:sigma2}
\begin{split}
\dfrac{3}{2} \, \sigma^{\prime\prime}(0) &=- (q_O-3) \, \left( \dfrac{3}{2} \, \dfrac{\sigma^\prime}{\mathcal{H}}\Bigg|_O \right) + \left( \dfrac{3}{2} \, \dfrac{\sigma^\prime}{\mathcal{H}}\Bigg|_O \right)^2 \, \left( \dfrac{k^3}{k^0} \right)^2_O =\\
&= (q_O - 3) \, \left( \dfrac{3}{2} \, \Omega_m \, \mathcal{H}_O \, \delta(0) \right) + \left( \dfrac{3}{2} \, \Omega_m \, \mathcal{H}_O \, \delta(0) \right)^2 \, \left( \dfrac{k^3}{k^0} \right)_O^2
\end{split}
\end{equation}
where in the second line we used \eqref{eq:rel1}.

We also report, without proving them, two relations involving the first and second time derivatives of the Hubble parameter and the deceleration and the jerk parameters needed in our calculations:\footnote{See also \cite{capo}, \cite{pedi} and \cite{cinesi}.}
\begin{align}
\label{eq:H1}
\dot{\mathcal{H}} &= - \mathcal{H}^2 \, \left( 1 + q \right);\\
\label{eq:H2}
\ddot{\mathcal{H}} &= \mathcal{H}^3 \, \left( j + 3 \, q +2 \right).
\end{align}

\section{Christoffel symbols} \label{app:cs}

In this appendix we report the non-null Christoffel symbols for the Goode-Wainwright representation for both families needed for calculating the geodesics.

\paragraph{Family I}

\begin{subequations}
\begin{equation*}
\Gamma_{11}^0 = \Gamma_{22}^0 = \mathcal{H} \, \left( S^2 e^{2\nu} \right)
\end{equation*}
\begin{equation*}
\Gamma_{33}^0 = \left( \dfrac{\dot{H}}{H} + \dfrac{\dot{S}}{S} \right) \, \left( S^2W^2H^2 \right)
\end{equation*}
\begin{equation*}
\Gamma^1_{10} = \Gamma_{20}^2= \mathcal{H}
\end{equation*}
\begin{equation*}
\Gamma_{30}^3 =  \dfrac{\dot{H}}{H} + \dfrac{\dot{S}}{S} 
\end{equation*}
\begin{equation*}
\Gamma_{12}^1 = \Gamma_{22}^2 = - \Gamma^2_{11} = \dfrac{\partial}{\partial y} \ln{\left( e^\nu S \right)}
\end{equation*}
\begin{equation*}
\Gamma_{13}^1 = \Gamma_{23}^2 = \dfrac{\partial}{\partial r} \ln{\left( e^\nu S \right)}
\end{equation*}
\begin{equation*}
\Gamma_{11}^1 = \Gamma_{21}^2 = - \Gamma_{22}^1 = \dfrac{\partial}{\partial x} \ln{\left( e^\nu S \right)}
\end{equation*}
\begin{equation*}
\Gamma_{33}^1 = - \dfrac{W^2H^2}{e^{2\nu}} \, \dfrac{\partial}{\partial x} \ln{\left( HWS \right)}
\end{equation*}
\begin{equation*}
\Gamma_{33}^2 = - \dfrac{H^2W^2}{e^{2\nu}} \, \dfrac{\partial}{\partial y} \ln{\left( H \right)}
\end{equation*}
\begin{equation*}
\Gamma_{11}^3 = \Gamma_{22}^3 = - \dfrac{e^{2\nu}}{H^2W^2} \, \dfrac{\partial}{\partial r} \ln{\left( Se^\nu \right)}
\end{equation*}
\begin{equation*}
\Gamma_{31}^3 = \dfrac{\partial}{\partial x} \ln{\left( H \right)}
\end{equation*}
\begin{equation*}
\Gamma_{32}^3 = \dfrac{\partial}{\partial y} \ln{\left( H\right)}
\end{equation*}
\begin{equation*}
\Gamma_{33}^3 = \dfrac{\partial}{\partial r} \ln{\left( HWS \right)}
\end{equation*}
\end{subequations}

\paragraph{Family II}

\begin{subequations}
\begin{equation*}
\Gamma^0_{11} = \Gamma^0_{22} = \mathcal{H} (S^2e^{2\nu});
\end{equation*}
\begin{equation*}
\Gamma^0_{33} =  \left( H \, S \right)^2 \, \left( \dfrac{\dot{H}}{H} + \dfrac{\dot{S}}{S} \right);
\end{equation*}
\begin{equation*}
\Gamma^3_{03} = \left( \dfrac{\dot{H}}{H} + \dfrac{\dot{S}}{S} \right);
\end{equation*}
\begin{equation*}
\Gamma^1_{01} = \Gamma^2_{02} = \mathcal{H};
\end{equation*}
\begin{equation*}
\Gamma^1_{11} = \Gamma^2_{12} = -\Gamma^1_{22} =  \dfrac{\partial \ln\left({e^\nu}\right)}{\partial x};
\end{equation*}
\begin{equation*}
\Gamma^2_{22} = \Gamma^1_{12} = - \Gamma^2_{11} = \dfrac{\partial \ln\left({e^\nu}\right)}{\partial y};
\end{equation*}
\begin{equation*}
\Gamma^1_{33} = - \dfrac{H^2}{e^{2\nu}} \, \dfrac{\partial \ln H}{\partial x};
\end{equation*}
\begin{equation*}
\Gamma^2_{33} = - \dfrac{H^2}{e^{2\nu}} \, \dfrac{\partial \ln H}{\partial y};
\end{equation*}
\begin{equation*}
\Gamma^3_{13} = \dfrac{\partial \ln H}{\partial x};
\end{equation*}
\begin{equation*}
\Gamma^3_{23} = \dfrac{\partial \ln H}{\partial y};
\end{equation*}
\begin{equation*}
\Gamma^3_{33} = \dfrac{\partial \ln H}{\partial r}.
\end{equation*}
\end{subequations}

\newpage
\section{Null geodesics and redshift} \label{sec:geod}

In this appendix we report null geodesics for both families calculated using Christoffel symbols given in the previous appendix and the null condition given in section \ref{sec:null}.

\paragraph{Family I}
\begin{align}
\label{eq:k0fam1}
\dfrac{d \, k^0}{d \lambda} &= - \mathcal{H}\, (k^0)^2 + \dfrac{3}{2} \, \sigma \, (HWS\, k^3)^2;\\
\dfrac{d}{d\lambda}\, \left[ k^1(e^\nu\,S)^2 \right] &= \dfrac{1}{2} \, \dfrac{\partial}{\partial x}{(e^{2\nu}S^2)} \, \left[ \left(k^0\right)^2 - (HWS)^2 \, (k^3)^2\right] - e^{2\nu}S^2\, \left( k^3 \right)^2 \, \dfrac{\partial}{\partial x} \, \ln{\left( {H} \right)}; \\
\dfrac{d}{d\lambda}\, \left[ k^2\, (e^\nu\,S)^2 \right] &= \dfrac{1}{2} \, \dfrac{\partial}{\partial y}{(e^{2\nu}S^2)} \, \left[ \left(k^0\right)^2 - (HWS)^2 \, (k^3)^2\right] - e^{2\nu}S^2\, \left( k^3 \right)^2 \, \dfrac{\partial}{\partial y} \, \ln{\left( {H} \right)}; \\
\label{eq:k3fam1}
\dfrac{d}{d\lambda} \, \left[ k^3\, (HSW)^2 \right] &= (HWS)^2 \, \left[ \left( k^0\right)^2 - \left(k^3\right)^2 \, \left( HWS \right)^2\right]\, \dfrac{\partial}{\partial r}\, \ln\left[(e^\nu\,S)\right] + \\ \nonumber
&+ \dfrac{1}{2} \,\dfrac{\partial}{\partial r} \,  \left[HWS\right]^2 \, (k^3)^2.
\end{align}
\paragraph{Family II}
\begin{align}
\label{eq:k0fam2}
\dfrac{d \, k^0}{d \lambda} &= - \dfrac{d}{d\lambda} \, \ln(S) \, k^0 + \dfrac{3}{2} \, \sigma \, (HS\, k^3)^2 = - \mathcal{H}\, (k^0)^2 + \dfrac{3}{2} \, \sigma \, (HS\, k^3)^2;\\\nonumber
\\
\dfrac{d}{d\lambda} \left[ k^0 \, S \right]  &= \dfrac{3}{2} \, \sigma \, \left( HSk^3 \right)^2;\\
\dfrac{d}{d \lambda}\, \left[ k^1\,(e^\nu\, S)^2 \right] & =\dfrac{1}{2} \, \dfrac{\partial}{\partial x}{(e^{2\nu}S^2)} \, \left[ \left(k^0\right)^2 - (HS)^2 \, (k^3)^2\right] - e^{2\nu}S^2\, \left( k^3 \right)^2 \, \dfrac{\partial}{\partial x} \, \ln{\left( {H} \right)}; \\
\dfrac{d}{d \lambda}\, \left[ k^2\,(e^\nu\, S)^2 \right] &= \dfrac{1}{2} \, \dfrac{\partial}{\partial y}{(e^{2\nu}S^2)} \, \left[ \left(k^0\right)^2 - (HS)^2 \, (k^3)^2\right] - e^{2\nu}S^2\, \left( k^3 \right)^2 \, \dfrac{\partial}{\partial y} \, \ln{\left( {H} \right)}; \\
\label{eq:k3fam2}
\dfrac{d}{d \lambda}\, \left[ k^3 \, (HS)^2 \right] &= \dfrac{1}{2} \, (k^3)^2 \, \dfrac{\partial}{\partial r} \, H^2.
\end{align}

\acknowledgments

The author wishes to thank K. Bolejko and M.N.C\`el\`erier and the anonymous referee for their patience and helpful suggestions. 

\paragraph{Note added} The solutions to the Szekeres model with a csmological source were first found by J.Barrow and J.S.Schabes in \cite{bss}. I am sorry for my omission.

\end{document}